\documentclass[11pt,twoside]{article}
\usepackage{asp2010}
\usepackage{natbib}

\resetcounters

\begin{document}

\title{Oscillations and magnetic fields in the G8 star EK Eridani}
\author{T. H. Dall$^1$, M. Cunha$^2$, K. G. Strassmeier$^3$, D. Stello$^4$, and H. Bruntt$^5$
\affil{$^1$European Southern Observatory, Karl-Schwarzschild-Str. 2, 85748 Garching bei M\"unchen, Germany}
\affil{$^2$Centro de Astrofísica da Universidade do Porto, Rua das Estrelas, 4150 Porto, Portugal}
\affil{$^3$Astrophysical Institute Potsdam, An der Sternwarte 16, 14482 Potsdam, Germany}
\affil{$^4$Sydney Institute for Astronomy, School of Physics, The University of Sydney, Sydney, 2006 NSW, Australia}
\affil{$^5$Institute for Physics and Astronomy, Aarhus University, Ny Munkegade, 8000 Aarhus C, Denmark}
}

\begin{abstract}
Asteroseismology can provide information that is otherwise not easily accessible, like the stellar mass and the evolutionary stage. Strong magnetic fields are usually accompanied by rapid rotation, which makes asteroseismology difficult due to spectral line broadening. We have found what may turn out to be the Rosetta Stone of the stars: A slowly rotating solar-like star with a strong magnetic field. We have recently detected solar-like oscillations in this active sub-giant, but with amplitudes much lower than expected. We suggest that the large-scale magnetic field alters the pulsations, which become magnetoacoustic in nature. Here we present our results and discuss possible implications and how this may open up a new frontier in the studies of magnetic fields and stellar evolution.
\end{abstract}

\section{Introduction}
EK Eri is a unique case of a slowly rotating ($v \sin i < 1$~km/s) G8 sub-giant, exhibiting brightness variations with a period of 308 days due to star spots being rotated across the projected surface \citep{strassmeier+1999}. It is significantly over-active with respect to its rotation rate and evolutionary state, deviating severely from the normal period-activity relation. Its magnetic field is large scale and dominated by a polodial mostly axisymmetric component, resembling a dipole with a surface strength of approximately 270~G \citep[][see also Auriere et al., these proceedings]{auriere+2008}.

The generally accepted explanation of the period-activity discrepancy is that EK Eri is the descendant of a magnetic Ap star \citep{stepien1993}. Assuming conservation of magnetic flux, EK Eri on the ZAMS would have had a field strength of a few kG, which is well within the typical range for magnetic Ap stars. However, this simple assumption is almost certainly inadequate since, as the star evolves off the main sequence, the growing convection zone interacts with the original field. Rather, a combination of fossil field acting as the seed for a growing dynamo-generated field is required, but
the model calculations are severely hampered by the lack of reliable mass and radius estimates for the star. 

\section{Astroseismic results}
We have recently \citep{dall+2010} detected oscillations in EK Eri based on three half-nights of HARPS data. This is the first detection of solar-like oscillations in an active giant star. Due to the short time coverage we could not extract the individual oscillation frequencies nor the large separation, $\Delta\nu$, between modes of successive radial order. However, we were able to estimate the amplitude per mode from the excess power in the Fourier spectrum using the approach of \citet{kjeldsen+2005}, and we find, at the position of maximum power $\nu_\mathrm{max} = 320 \pm 32$~$\mu$Hz, that the peak amplitude per radial mode is $\simeq 0.15$~ms$^{-1}$ . This amplitude is at least a factor of 3 lower than expected based
on the scaling relations of \citet{kjeldsen+2008}.

We suggest that the dominant cause of this discrepancy is due to the magnetic field.
A large scale magnetic field  will influence the pulsations in the atmosphere of the star, which will become magnetoacoustic in nature, such as is known to happen in sunspots and in main-sequence Ap stars. The pulsation velocity field will acquire a significant horizontal component, which will depend strongly on the inclination of the magnetic field. Thus, not only the extracted radial velocity amplitude will be different from what would be expected in the absence of the field, but it will depend on the relation between the position of the observer, the magnetic axis, and the pulsation axis. In the following Section we will discuss this effect.

\section{Magnetic effect on acoustic oscillations}
Studies of oscillations in main-sequence magnetic pulsators --- the roAp stars --- show that in the presence of a large-scale magnetic field that permeates the surface of the star the pulsational velocity is severely modified when compared to the non-magnetic case \citep[e.g.,][]{dziembowski+goode1996,cunha+gough2000,saio+gautschy2004}.

In particular, in the outer layers of the star the velocity associated with low degree modes acquires a significant horizontal component. Also, the relative size of the radial and horizontal components depends critically on the inclination of the field. Thus, not only the extracted radial velocity amplitude will be different from what may be expected in the absence of the field, but it will depend on the relation between the position of the observer, the magnetic axis, and the pulsation axis.

Moreover, the latitudinal dependence of the mode amplitude is strongly modified. The simplest example of this is the case of radial modes, which amplitudes at the surface cease to be spherically symmetric due to the direct effect of the magnetic field \citep[see e.g.,][Fig. 18]{saio+gautschy2004}.

Acoustic pulsations are modified in the presence of a magnetic field. If the magnetic field permeates the surface of the star, there will always be a region where the Lorentz
force and the gas-pressure gradient acting on a perturbed element of fluid are comparable. In that region the effect of the magnetic field on the dynamics of the oscillations cannot be regarded as a small perturbation. 

\begin{figure}[!ht]
\plotone{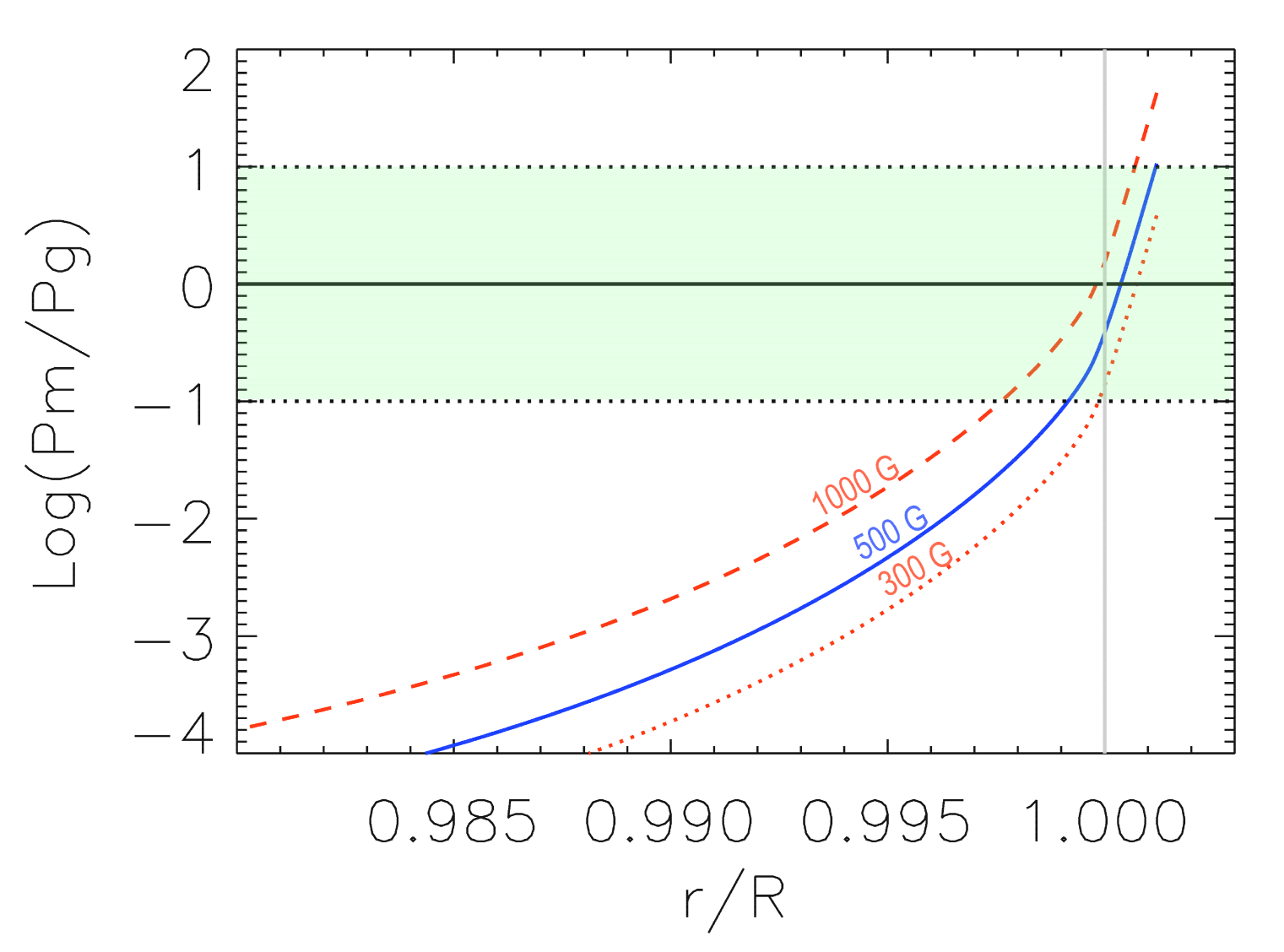}
\caption{\label{fig1}
A model for the magnetic boundary layer in EK Eri. 
The model has $M = 2M_\odot$, $T_\mathrm{eff} = 5100$~K and $\log(L/L_\odot) = 1.2$.  
The region where the gas and magnetic pressures (Pg and Pm, respectively) are comparable is shown for three different values for the magnetic field: 300 G (dotted red line), 500 G (full blue line), 1000 G (red dashed line). The shaded area between the two horizontal dotted lines show the upper and lower limits of the region where the pressures are comparable, while the full horizontal line marks where the two are the same. The gray vertical line shows the position of the photosphere. For B = 300 G, the region of interest is almost entirely above the photosphere, while for B = 500 G and higher, the gas and magnetic pressures are similar already in the photosphere. }
\end{figure}

The extent and position of the singular region --- or magnetic boundary layer --- depends on the properties of the outer layers of the star, as well as on the magnetic field intensity.  In Fig.~\ref{fig1} we show the position of that region for EK Eri, estimated as the region where gas and magnetic pressures are comparable.

According to \citet{auriere+2008}, the mean surface magnetic field of EK Eri is 270~G. The value of maximum field at the surface cannot be deduced from the mean surface field alone, but one expects that for a simple dipolar geometry it will be only slightly larger. 

For a magnetic field magnitude of 300~G, we see from Fig.~\ref{fig1} that the region where the magnetic field has a direct effect on the dynamics lies mostly above the photosphere. For oscillations of frequency below the acoustic cutoff frequency, the greatest direct impact of the field will thus take place outside the cavity of the oscillations, in the region where the latter are evanescent. On the other hand, for a magnetic field of 500~G, the region of direct magnetic field influence extends all the way to the photosphere and is expected to result in stronger perturbations both to the eigenfrequencies and eigenfunctions. 

One of the consequences of the surface velocity being no longer radial, in the presence of a magnetic field, is that for the same underlying amplitude, the observed amplitude at the photosphere will be different from the non-magnetic case. That could well be the origin of the excessively low amplitude per radial mode observed in EK Eri.  

To test that possibility, and attempt to build additional observational test, we are planning a detailed study of the pulsations of EK Eri including the effect of the magnetic field. Along with magnetic models, such a study requires improved asteroseismic data and additional information about the star's magnetic field.

\section{Conclusion}

We have argued that the solar-like oscillations we have detected in EK Eri are of magneto-acoustic nature, and hence potentially contains information on the magnetic field and its structure as well as on the internal structure of the stellar matter.  Based on a simple model we predict the maximum surface magnetic field to be at least 300~G, which  is consistent with recent measurements. More detailed models will be developed for further study.

For the first time we now have the opportunity to study the magnetic field and the internal structure of a star other than the Sun without the complications introduced by rapid rotation.  Like a Rosetta Stone, EK Eri may lead us to a deeper understanding of magnetic fields and solar-like oscillations, and more importantly, to a new understanding of the interplay between the oscillations and the magnetic field.


\acknowledgements The authors would like to thank M.~Auri\`ere and P.~Petit for good discussions and helpful suggestions.

\bibliographystyle{bibtex/natbib}
\bibliography{dall_t}

\begin{thebibliography}{}
\expandafter\ifx\csname natexlab\endcsname\relax\def\natexlab#1{#1}\fi
\expandafter\ifx\csname url\endcsname\relax
  \def\url#1{\texttt{#1}}\fi
\expandafter\ifx\csname urlprefix\endcsname\relax\def\urlprefix{URL }\fi
\providecommand{\eprint}[2][]{\url{#2}}

\bibitem[{{Auri{\`e}re} et~al.(2008){Auri{\`e}re}, {Konstantinova-Antova},
  {Petit}, {Charbonnel}, {Dintrans}, {Ligni{\`e}res}, {Roudier}, {Alecian},
  {Donati}, {Landstreet}, \& {Wade}}]{auriere+2008}
{Auri{\`e}re}, M., {Konstantinova-Antova}, R., {Petit}, P., {Charbonnel}, C.,
  {Dintrans}, B., {Ligni{\`e}res}, F., {Roudier}, T., {Alecian}, E., {Donati},
  J.~F., {Landstreet}, J.~D., \& {Wade}, G.~A. 2008, \aap, 491, 499.
  \eprint{0810.2228}

\bibitem[{{Cunha} \& {Gough}(2000)}]{cunha+gough2000}
{Cunha}, M.~S., \& {Gough}, D. 2000, \mnras, 319, 1020

\bibitem[{{Dall} et~al.(2010){Dall}, {Bruntt}, {Stello}, \&
  {Strassmeier}}]{dall+2010}
{Dall}, T.~H., {Bruntt}, H., {Stello}, D., \& {Strassmeier}, K.~G. 2010, \aap,
  514, A25+. \eprint{1003.0433}

\bibitem[{{Dziembowski} \& {Goode}(1996)}]{dziembowski+goode1996}
{Dziembowski}, W.~A., \& {Goode}, P.~R. 1996, \apj, 458, 338

\bibitem[{{Kjeldsen} et~al.(2008){Kjeldsen}, {Bedding}, {Arentoft}, {Butler},
  {Dall}, {Karoff}, {Kiss}, {Tinney}, \& {Chaplin}}]{kjeldsen+2008}
{Kjeldsen}, H., {Bedding}, T.~R., {Arentoft}, T., {Butler}, R.~P., {Dall},
  T.~H., {Karoff}, C., {Kiss}, L.~L., {Tinney}, C.~G., \& {Chaplin}, W.~J.
  2008, \apj, 682, 1370. \eprint{0804.1182}

\bibitem[{{Kjeldsen} et~al.(2005){Kjeldsen}, {Bedding}, {Butler},
  {Christensen-Dalsgaard}, {Kiss}, {McCarthy}, {Marcy}, {Tinney}, \&
  {Wright}}]{kjeldsen+2005}
{Kjeldsen}, H., {Bedding}, T.~R., {Butler}, R.~P., {Christensen-Dalsgaard}, J.,
  {Kiss}, L.~L., {McCarthy}, C., {Marcy}, G.~W., {Tinney}, C.~G., \& {Wright},
  J.~T. 2005, \apj, 635, 1281

\bibitem[{{Saio} \& {Gautschy}(2004)}]{saio+gautschy2004}
{Saio}, H., \& {Gautschy}, A. 2004, \mnras, 350, 485

\bibitem[{{St\c{e}pie\'{n}}(1993)}]{stepien1993}
{St\c{e}pie\'{n}}, K. 1993, \apj, 416, 368

\bibitem[{{Strassmeier} et~al.(1999){Strassmeier}, {St{\c e}pie{\' n} },
  {Henry}, \& {Hall}}]{strassmeier+1999}
{Strassmeier}, K.~G., {St{\c e}pie{\' n} }, K., {Henry}, G.~W., \& {Hall},
  D.~S. 1999, \aap, 343, 175

\end{thebibliography}

\end{document}